\documentclass[%
 reprint,
 amsmath,amssymb,
 aps,
pre,
floatfix,
]{revtex4-2}

\usepackage[utf8]{inputenc}
\usepackage[T1]{fontenc}
\usepackage{amsmath}
\usepackage{amsfonts}
\usepackage{amssymb}
\usepackage[version=4]{mhchem}
\usepackage{stmaryrd}
\usepackage{color}

\usepackage{graphicx}

\newcommand{\phd}{P_{\parallel}}
\newcommand{\piso}{P_\text{iso}}
\newcommand{\riso}{\rho_\text{c,iso}}

\newcommand{\kt}{k_\text{B}T}

\newcommand{\rmax}{\rho_\text{max}}

\usepackage[normalem]{ulem}

\begin{document}

\title{{Competition between shape anisotropy and deformation in the ordering and close packing properties of quasi-one-dimensional hard superellipse fluids}}

\author{Sakineh Mizani}
\affiliation{Institute for Applied Physics, University of Tübingen, Auf der Morgenstelle 10, 72076 Tübingen, Germany}
\author{Péter Gurin}
\affiliation{Physics Department, Centre for Natural Sciences, University of Pannonia, PO Box 158, Veszprém, H-8201 Hungary}
\author{Szabolcs Varga}
\affiliation{Physics Department, Centre for Natural Sciences, University of Pannonia, PO Box 158, Veszprém, H-8201 Hungary}
\author{Martin Oettel}
\affiliation{Institute for Applied Physics, University of Tübingen, Auf der Morgenstelle 10, 72076 Tübingen, Germany}

\date{\today}

\begin{abstract}
{
We investigate the orientational ordering and close packing behavior
of a quasi-one-dimensional (q1D) system of hard superellipses, where
the centers of the particles are confined to a line while they can
rotate freely within a two-dimensional plane. The particle shape is
tuned between an ellipse and a rectangle by varying the deformation
parameter ($n$). The elongation of the particle is changed with the
help of the aspect ratio ($k$). The pressure ratio between freely
rotating and parallel hard superellipses, which displays a single
peak, serves as an effective marker for the continuous structural
change from quasi-isotropic to nematic ordering. Our findings reveal a
competition between the parameters $k$ and $n$, with $k$ promoting
nematic alignment and $n$ favoring tetratic ordering. Notably, in the
close-packing regime, the packing properties become independent of
$k$, as the relevant exponents depend solely on $n$. Furthermore,
certain combinations of these exponents exhibit universality,
remaining invariant with respect to particle shape.
}
\end{abstract}

\maketitle

\section{Introduction}
The study of liquid crystalline phases and their transitions has been
a subject of extensive research, with particular focus on the
isotropic-nematic (IN) transition in three-dimensional
systems \cite{Parsons1979,EspositoEvans1994}. In this regard, hard
particle models have proven instrumental in understanding the
mechanisms driving these phase
transitions \cite{Onsager1942,Onsager1949,VroegeLekkerkerker1992,Kike-Yuri-Luis_JPhysCondMat_2014}. Among these, hard spherocylinders (HSC) comprising a
cylindrical rod of length $L$ and diameter $D$, capped at each end by
hemispheres of diameter $D$ and hard ellipsoids with $a$ and $b$
semiaxis lengths have emerged as paradigmatic
systems \cite{StroobantsLekkerkerkerFrenkel1986,StroobantsLekkerkerkerFrenkel1987,VeermanFrenkel1991}. The
appeal of these models lies in their simplicity while still capturing
the essential physics of real mesogens. For hard spherocylinders, the
phase behavior is remarkably rich, exhibiting isotropic ($I$), nematic
($N$), smectic-A ($SmA$), and solid ($K$) phases with increasing
density \cite{Frenkel1988,FrenkelLekkerkerkerStroobants1988,VeermanFrenkel1990}. The
stability of these phases strongly depends on the aspect ratio $(k=1+L
/ D)$, with the system reducing to hard spheres in the limit
$k \rightarrow 1$ and recovering the Onsager limit of infinitely long
rods as
$k \rightarrow \infty$ \cite{Stroobants1992,Stroobants1994}. Early
computer simulations by Frenkel and coworkers and subsequent studies
have established that for $k \geq 6$, the system displays a clear
sequence of $I-N-S m A-K$ transitions \cite{Frenkel1988}. However, for
intermediate aspect ratios around $k \approx 4.2$, the system exhibits
a direct isotropic to smectic-A transition, leading to the existence
of both $I-S m A-K$ and $I-N-S m A$ triple
points \cite{EspositoEvans1994,McGrotherWilliamsonJackson1996,BolhuisFrenkel1997}. In
contrast, hard ellipsoids, characterized by their aspect ratio $k$
show a simpler phase diagram that lacks the smectic-A phase observed
in HSCs \cite{FrenkelMulder1985,AllenWilson1989,AllenEvansFrenkelMulder1993,Gerardo_JChemPhys_2012}. The IN
transition only becomes stable for $k>5$, with the transition becoming
increasingly well-described by Onsager theory at higher aspect
ratios \cite{CampMasonAllenKhareKofke1996}. Theoretical approaches,
particularly the Parsons-Lee scaling of Onsager theory, have proven
remarkably successful in describing the IN transition, accurately
predicting both the transition pressures and coexisting densities,
though they somewhat less accurately capture the orientational
distribution functions \cite{Lee1987, Lee1988}.

Going to lower dimensions crucially affects the liquid crystalline
phase behavior. As systems are confined from three to two dimensions,
the competition between orientational entropy and excluded volume
effects is significantly altered by the elimination of one rotational
degree of freedom \cite{DamascenoEngelGlotzer2012,
AndersAhmedSmithEngelGlotzer2014,BolhuisFrenkel1997,AvendanoEscobedo2012,MizaniNaghaviVarga2023}
and the ordering transition of two-dimensional (2D) systems strongly
depends on the particle
shape \cite{QiGraafQiaoMarrasMannaDijkstra2012,FrenkelEppenga1985,ShahKangKohlstedtAhnGlotzerMonroeSolomon2012,DonevBurtonStillingerTorquato2006,NelsonHalperin1979,Young1979,BernardKrauth2011,
EngelAndersonGlotzerIsobeBernardKrauth2013, CuestaFrenkel1990}. For
instance, while hard ellipses exhibit nematic ordering at aspect
ratios of approximately 4 \cite{CuestaFrenkel1990}, 2D hard
spherocylinders require a significantly higher aspect ratio of about 8
to achieve similar ordering \cite{BatesFrenkel2000}.  As another
example, recent work on phases of hard superellipses highlights how
varying particle shape parameters (e.g. aspect ratio and deformation
parameter $n$) leads to a diverse range of phases, such as isotropic,
nematic, and various crystalline
forms \cite{TorresDiazHendleyMishraYehBevan2022}.  When comparing 2D
and 3D, one notices a significant enhancement of the critical aspect
ratio for the onset of nematic ordering $k \approx 8$ for 2D hard
spherocylinders vs. $k \approx 4.5$ for their 3D
counterparts \cite{BolhuisFrenkel1997,BatesFrenkel2000}.  Furthermore,
the order of the isotropic-nematic (IN) phase transition changes from
first-order to a continuous Kosterlitz-Thouless (KT)
transition \cite{FrenkelEppenga1985,mizani2020,HalperinNelson1978},
illustrating how dimensional reduction not only affects stability
thresholds but also alters the nature of phase transitions
\cite{BatesFrenkel2000,XuLiSunAn2013,KosterlitzThouless1973,li2021phase,NelsonHalperin1979-1,Young1979-1,BernardKrauth2011-1,Straley1971,ZhengHan2010}.

Further confinement of the translational degree of freedom to narrow
channels and eventually to one dimension while retaining orientational
freedom defines quasi-one-dimensional (q1D) geometries and brings
about new effects
\cite{GurinaMizaniVarga2024,MizaniOettelGurinaVarga2024}, as a
bridge between one- and two-dimensional
behavior \cite{BasurtoGurinVargaOdriozola2021,
JinWangChanZhong2021,LebowitzPercusTalbot1987,KantorKardar2009}. Although
these systems do not exhibit true thermodynamic phase transitions,
they inherit some properties of higher dimensional systems such as the
orientational ordering~\cite{MizaniOettelGurinaVarga2024}.  In
addition to this, the close packing behaviour of q1D systems is very
interesting, because the particle's shape has a great influence on the
character of orientational correlations. For example, hard rectangles
confined to move along a straight line exhibit a divergent
orientational correlation lengths, while hard ellipses do
not \cite{KantorKardar2009-1}.  In addition to this, the average
angular fluctuations exhibit a power-law dependence on pressure in the
vicinity of the close-packing density with different exponents for
rectangles and ellipses. Furthermore, the introduction of slight
transverse positional freedom can lead to complex phenomena such as
the emergence of glassy behavior and fragile-to-strong fluid
crossovers, emphasizing that subtle changes in confinement can
dramatically affect the phase
properties \cite{AshwinYamchiBowles2013,GodfreyMoore2014,
GodfreyMoore2015,RobinsonGodfreyMoore2016,HuCharbonneau2018,
ZhangGodfreyMoore2020,HuCharbonneau2021,MonteroSantos2023}.

The present work investigates the role of particle geometry (aspect
ratio and deformation parameter) in systems of q1D superellipses,
where particles have twofold symmetry that can be continuously
deformed between elliptical and rectangular shapes.  In our previous
work \cite{MizaniOettelGurinaVarga2024} we have shown that for q1D
superdisks (interpolating between circular disks and squares) the
continuous change from an quasi-isotropic to a tetratically ordered state is
accompanied with a characteristic maximum in the pressure ratio {
$P/\phd$ as a function of number density $\rho$, where $P$ is the
pressure of freely rotating hard superdisks and $\phd$ is that of
parallel superdisks.} This maximum suitably defines the location of
the quasi-isotropic--tetratic structural
change~\cite{MizaniOettelGurinaVarga2024}.  Here we show that the
pressure ratio maximum likewise defines the location of the
quasi-isotropic--nematic change, {furthermore, the aspect ratio
and the deformation parameter compete with each other, 
moving the $P/\phd$ maximum in opposite directions. } We examine
the aspect ratio dependence of characteristic exponents in
the close-packing limit, for which universal laws had been found in
the q1D superdisk system \cite{MizaniOettelGurinaVarga2024}.  It turns
out that the close-packing exponents are independent of aspect ratio and
thus the validity of the universal laws extends to the whole class of
superelliptic particles.

\begin{figure}[t]
    \centering
    \includegraphics[width = \columnwidth,trim=3cm 3cm 12cm 0.1cm,clip ]{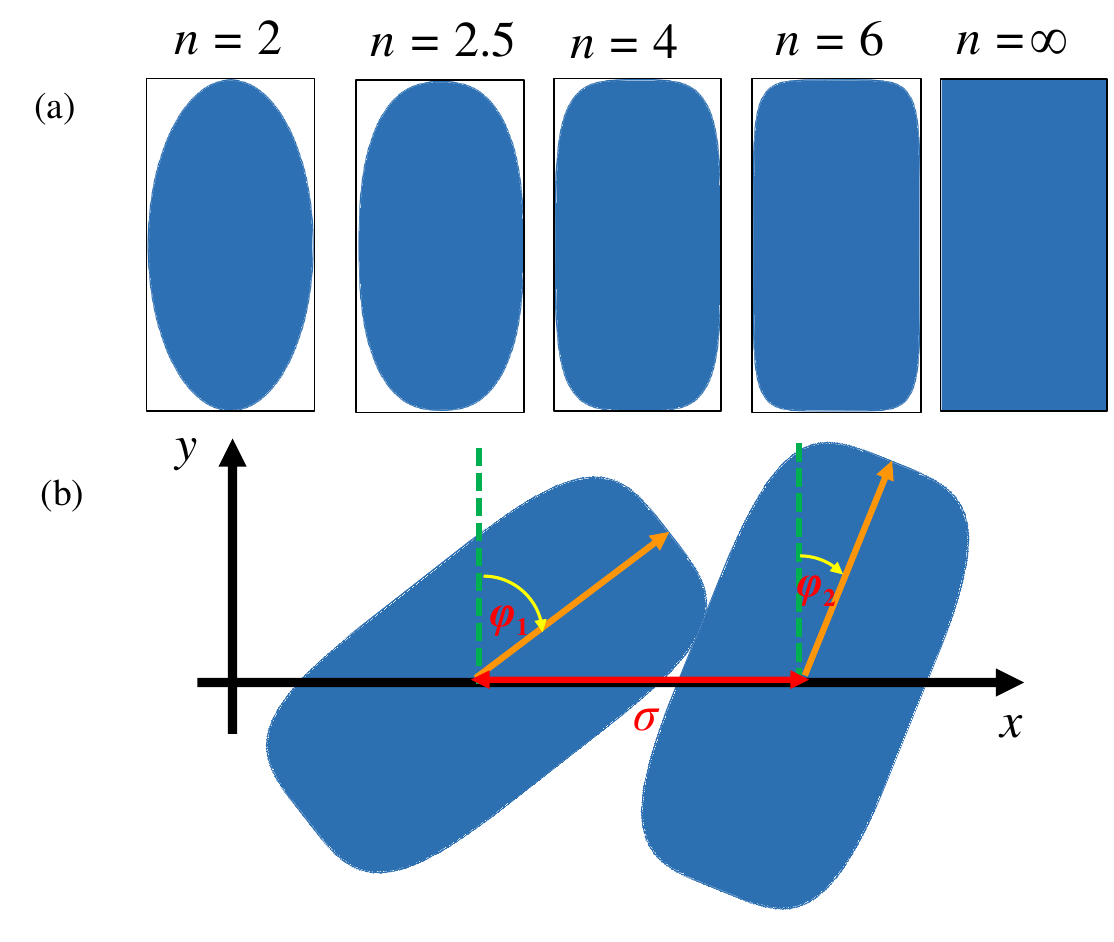}
	\caption{(a) The effect of the deformation
	parameter \textit{n} on the shape of hard superellipses with
	aspect ratio $k=2$ and (b) contact distance between two
	neighboring hard superellipses with orientations $\varphi_{1}$
	and $\varphi_{2}$ in the q1D geometry in which the centers of
	the superellipses are constrained to move along
	the \textit{x}-axis but can rotate freely within the $x$-$y$
	plane. } \label{fig:geometry}
\end{figure}

\section{Hard Superellipse Model}

In the q1D model, the centers of $N$ particles are constrained to move
along a straight line of length $L$, and they can freely rotate around
their center about an angle ($\varphi$) in the $x$--$y$ plane (see
Fig.~\ref{fig:geometry}).  The geometric boundary of a superellipse
particle, centered at the origin with orientation $\varphi=0$, is
given by
\begin{equation}
h:=\left|\frac{x}{a}\right|^{n}+\left|\frac{y}{b}\right|^{n}-1=0\;,
	\label{eq:boundary}
\end{equation}
The deformation parameter $n \in[2, \infty)$, provides a continuous
transformation between hard ellipses ($n=2$) and hard rectangles
$(n \rightarrow \infty)$.  The geometric parameters $a$ and $b$ define
the semi-minor and semi-major axes respectively. \{By choosing
$d=2a$ to be the unit of length, the aspect ratio ($k=b/a$) is
the free parameter, which can interpolate between the superdisk
($k=1$) and the needle ($k \rightarrow \infty$), i.e.
with increasing $k$ the particle becomes more elongated and with increasing $n$ more rectangular (see Fig.~\ref{fig:geometry}).
While both parameters $k \neq 1 $, $n \neq 2$  give rise to anisotropic shape, the 
difference between them is
that $n$ breaks the continuous rotational symmetry to be fourfold
$C_4$ symmetry, while $k$ gives rise to twofold $C_2$ rotational
symmetry. 
The combined effect of $n$ and $k$ is that the symmetry of
the particle is always $C_2$.
This can be also seen in the structure of q1D fluid of hard superellipses, where the ordering is nematic.

The hard-core interaction between particles entails a minimum
separation distance ($\sigma$) between adjacent particles which can be
computed as follows. For a pair of superellipses in contact (indexed
by $i=1,2$) with orientations $\varphi_{i} \in[-\pi/2,\pi/2]$ and
center position $x_1=0$ and $x_2=\sigma$ the boundary curves $h_i$ are
given by
\begin{eqnarray}
h_{i}&:=&\left|\frac{(x-x_i) \cos\varphi_{i} +y\sin\varphi_i}{a} \right|^{n}+ \nonumber \\
 & & \left|\frac{-(x-x_i) \sin\varphi_{i} +y\cos\varphi_i}{b} \right|^{n}-1=0\;. 
	\label{eq:hi}
\end{eqnarray}
The contact configuration is characterized by the antiparallel
alignment of the boundary gradients at the point of contact, expressed
as
\begin{equation}
\nabla h_{1}=-\mu \nabla h_{2} 
	\label{eq:gradient}
\end{equation}
with a positive constant $\mu$. The contact geometry is fully
determined by the coupled system of equations (\ref{eq:hi}) and
(\ref{eq:gradient}) with solutions for ($x, y, \sigma, \mu$) as
functions of the particle orientations $\varphi_{i}$ as well as $n$
and $k$.
{The important output of the solution of Eqs.~(\ref{eq:hi}) and
(\ref{eq:gradient}) is the contact distance, which depends on the
orientations of the neighboring particles,
i.e. $\sigma=\sigma(\varphi_1,\varphi_2)$. We use this quantity as an
input of the transfer operator method.}

\section{Theory}

Exact equilibrium properties are obtained with the transfer operator
method (TOM) which is particularly effective for systems characterized
by first-neighbor interactions, as it allows us to derive
thermodynamic and structural properties by solving an eigenvalue
problem
in NPT ensemble~\cite{LebowitzPercusTalbot1987}, where $N$ is the number of particles, $P$ is the pressure and $T$ is the temperature. The key quantities are the contact distance between two superellipses ($\sigma$) and the pressure, as the kernel of transfer operator is given by
\begin{eqnarray}
K\left(\varphi_{1}, \varphi_{2}\right) &=& \frac{\kt}{ P}\exp \left(-\frac{P}{k_\text{B} T} \sigma\left(\varphi_{1}, \varphi_{2}\right)  \right) ,
\end{eqnarray}
where $k_\text{B}$ is Boltzmann's constant. The eigenvalue equation of $K$ is given by
\begin{eqnarray}
	\label{eq:eigenvalue}
	\int_{-\pi / 2}^{\pi / 2} K\left(\varphi_{1}, \varphi_{2}\right) \psi_{i}\left(\varphi_{2}\right) d \varphi_{2}&=&\lambda_{i} \psi_{i}\left(\varphi_{1}\right) \;.
\end{eqnarray}
The solutions of
Eq.~(\ref{eq:eigenvalue}) yield the eigenvalues $\lambda_{i}$ and the
corresponding eigenfunctions $\psi_{i}(\varphi)$ which are normalized
by $\int_{-\pi / 2}^{\pi / 2} \psi_{i}^{2}(\varphi) d \varphi=1$.
From the largest eigenvalue $\lambda_{0}$ the (inverse) equation of
state $\rho(P)$ (where $\rho$ is the particle number density) is
obtained by
\begin{eqnarray}
\rho =-\frac{\lambda_{0}}{\frac{d \lambda_{0}}{d\left(\beta P \right)}} \;.
\label{eq:rho}
\end{eqnarray}

For $k=1$ and $n=2$, we have a system of q1D circular disks and there
is no angular dependence to be accounted for in Eq.~(\ref{eq:eigenvalue}).
{This hard disk system is equivalent to the orientationally frozen
system of hard superellipses if all particles are parallel. For
$\sigma=d$, the solution of Eqs.~(\ref{eq:eigenvalue}) and
(\ref{eq:rho}) becomes identical with the well-known equation of state
of hard rods derived by Tonks~\cite{Tonks1936},}
\begin{equation}
\phd=\frac{\kt \rho}{1-\rho d} \;.
	\label{eq:ptonks}
\end{equation}
{Note that we set $d=2a$ because the superellipses become parallel with
their shortest length along the confining line in the close-packing
limit.  The first close-packing exponent measures the contribution of the
orientational fluctuations to the pressure in the close-packing limit
as it is defined by}
\begin{equation}
 \alpha=\lim _{\rho d \rightarrow 1} \frac{P}{\phd} \, .
 \label{eq:alpha}
\end{equation} 
The normalized orientational distribution function (ODF) is given by
\begin{equation}
f(\varphi)=\psi_{0}^{2}(\varphi)\;,
	\label{eq:odf}
\end{equation}
which in the isotropic state is given by $f=1 / \pi$ while a peak at $\varphi=0$ indicates nematic order. 
A corresponding nematic order parameter is defined by $S=\int_{-\pi / 2}^{\pi / 2} f(\varphi) \cos (2 \varphi) d \varphi$, entailing $0<S<1$, i.e. $S=0$ for an isotropic state and
$S=1$ for a perfectly ordered nematic state. The average angular fluctuations are given by
\begin{equation}
\left\langle\varphi^{2}\right\rangle=\int_{-\pi / 2}^{\pi / 2} f(\varphi) \varphi^{2} d \varphi \;,
 \label{eq:angfluc}
\end{equation}
which permits the definition of our second close-packing exponent ($ \langle\varphi^{2}\rangle \propto P^\beta$)
\begin{equation}
	\beta=\lim _{P \rightarrow \infty} \frac{d \ln \left\langle\varphi^{2}\right\rangle}{d \ln P}\;.
 \label{eq:beta}
\end{equation}
If the particles along the line are numbered as $1 ...i+1$, the
orientational correlation function is defined as
\begin{equation}
	g_{2}(i)=\left \langle\cos (2(\varphi_1-\varphi_{i+1}))\right \rangle-S^{2} \propto \exp (-i / \xi)\;,
  \label{eq:g2}
\end{equation}
where the orientational correlation length $\xi$ is obtained from the
largest ($\lambda_0$) and the second largest ($\lambda_1$) eigenvalues
as follows
\begin{equation}
1 / \xi=\ln \left(\lambda_{0} / \lambda_{1}\right) \;. 
	\label{eq:xi}
\end{equation}
Therefore we define our third close-packing exponent ($ \xi \propto
P^\gamma$) as follows
\begin{equation}
	\gamma=\lim _{P \rightarrow \infty} \frac{d \ln \xi}{d \ln P} \;.
 \label{eq:gamma}
\end{equation}
For the numerical solution of Eq.~(\ref{eq:eigenvalue}), a trapezoidal
rule for the angular integration with a spacing $\Delta \varphi=\pi /
10^4$ is employed. The eigenvalues and eigenfunctions are obtained
using a successive iteration method, starting with an initial guess of
$\psi_{i}(\varphi)=\frac{1}{\sqrt{\pi}}$ for all systems.

\section{Results}

{
We start this section with showing the effect of deformation parameter
($n$) and aspect ratio ($k$) on the orientational distribution
function, which is obtained from Eqs.~(\ref{eq:eigenvalue}) and
(\ref{eq:odf}). 
\begin{figure}[ht]
        \includegraphics[width = 0.95\columnwidth,trim=1cm 0.5cm 3cm 1cm,clip ]{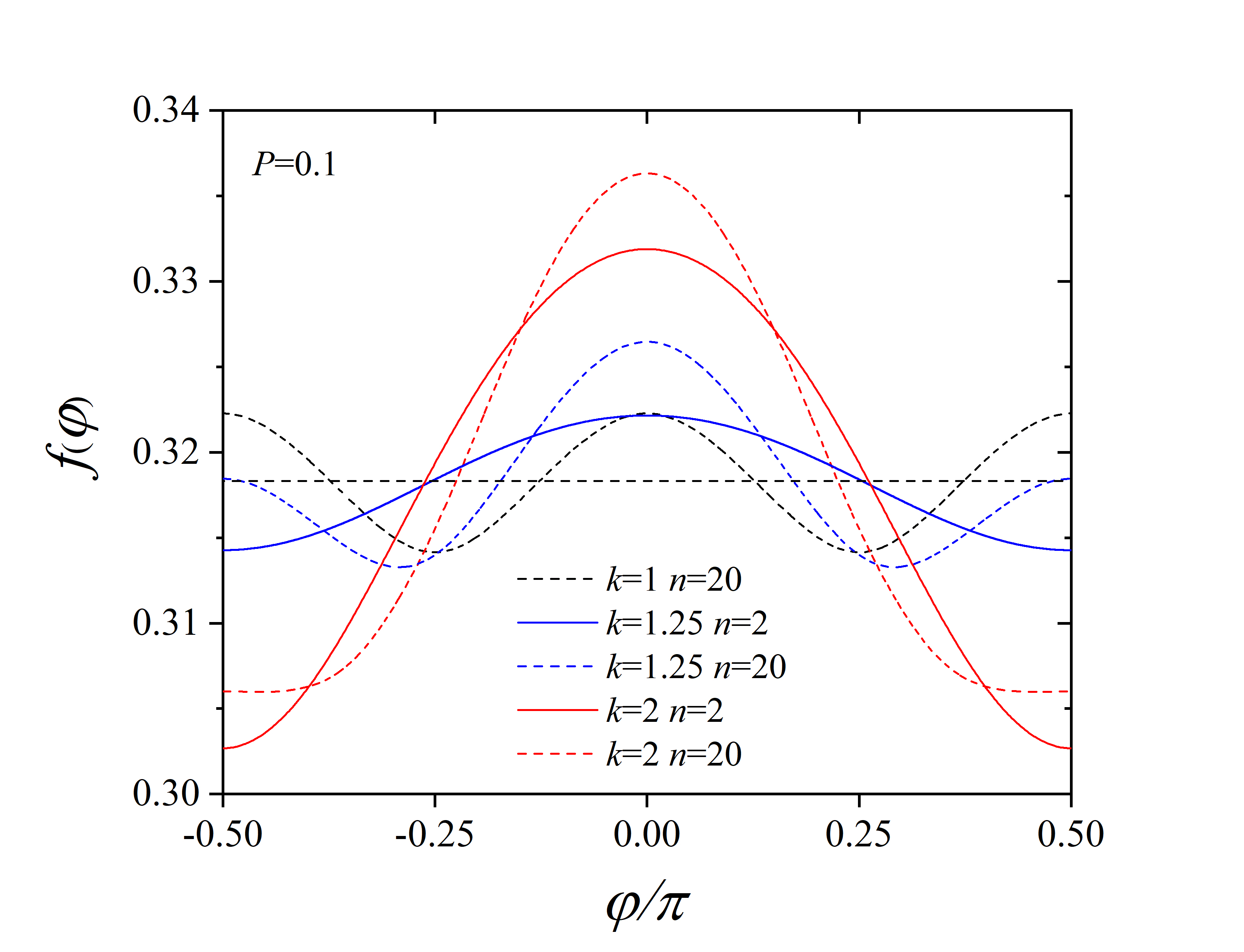}
        \includegraphics[width = 0.95\columnwidth,trim=1cm 0.5cm 3cm 1cm,clip ]{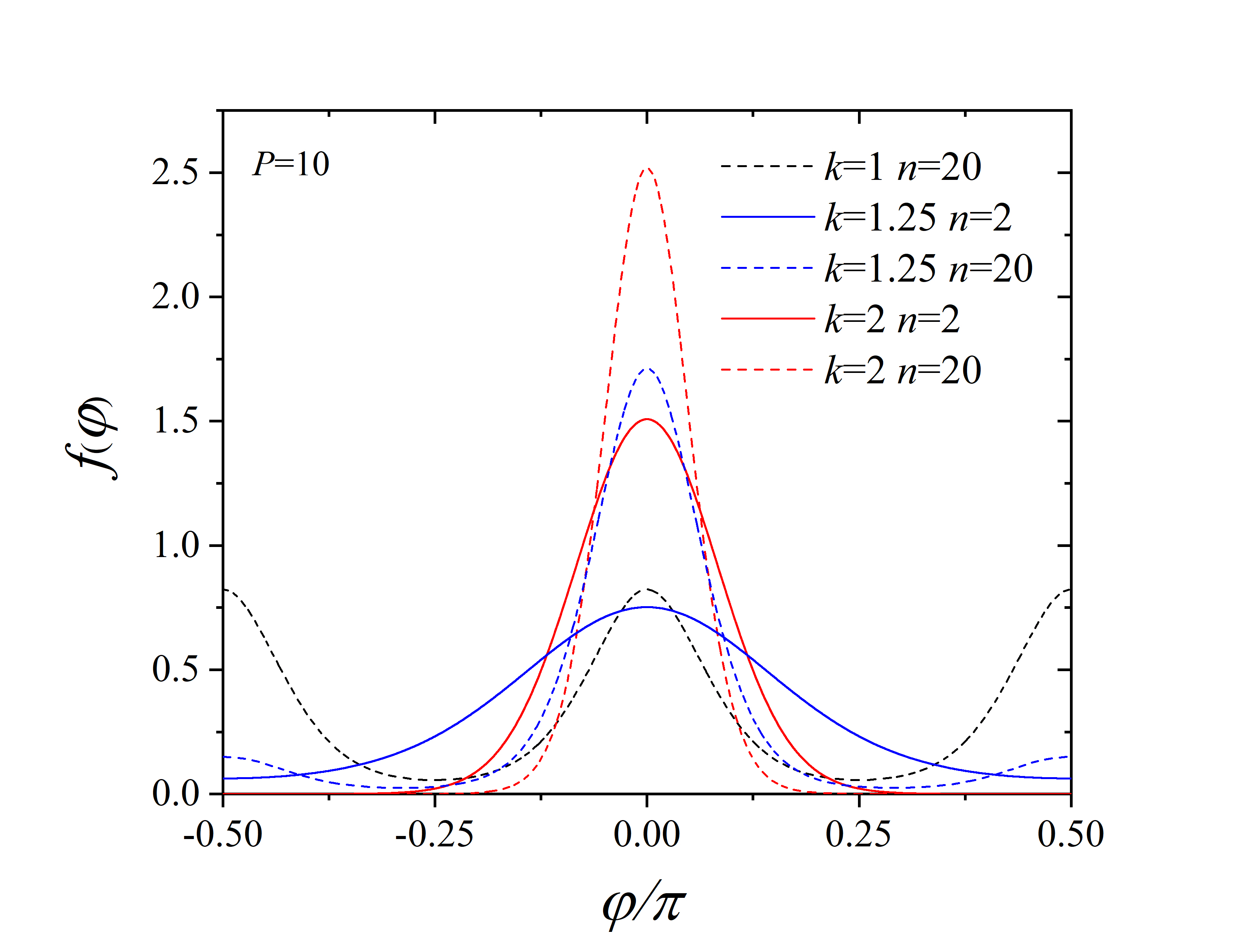}
	\caption{Effect of deformation parameter ($n$) and aspect
        ratio ($k$) on the orientation distribution function. The upper
	panel shows the results at $P^*=0.1$, while the lower one at
	$P^*=10$. The horizontal dashed line corresponds to the
	isotropic distribution ($f=1/\pi$).
    }  \label{fig:ODF}
\end{figure}
For $n>2$ or $k>1$, even at very low pressures it deviates from the isotropic solution $f=1/\pi$, see Fig.~\ref{fig:ODF}.
The tetratic structure,
for which in particular $f(0)=f(\pm\pi/2)$ holds,  exists only for $n>2$ and $k=1$ at
both low and high pressures. The case of $n=2$ and $k>1$ results in a
$f(\varphi)$, which has a maximum at $\varphi=0$ and a minimum at
$\varphi=\pm\pi/2$, corresponding to typical nematic order. In
the case of high $n$ and $k$ close to 1, we can observe an
intermediate orientational distribution between nematic and
tetratic ordering as a local maximum survives at
$\varphi=\pm\pi/2$. We can also see in Fig.~\ref{fig:ODF} that 
increasing $k$ completely destroys a remaining tetratic ordering
at both low and high pressures, e.g. for $n=20$ the local maximum disappears already at
$P^*=0.1$ for $k=2$. From these results it can be concluded that $k$
favors nematic, while $n$ tetratic ordering. 
In the following we examine how the competition between these two parameters influences
the thermodynamic properties.  }

\begin{figure}[t!]
        \includegraphics[width = 0.94\columnwidth,trim=1cm 0.5cm 3cm 2cm,clip ]{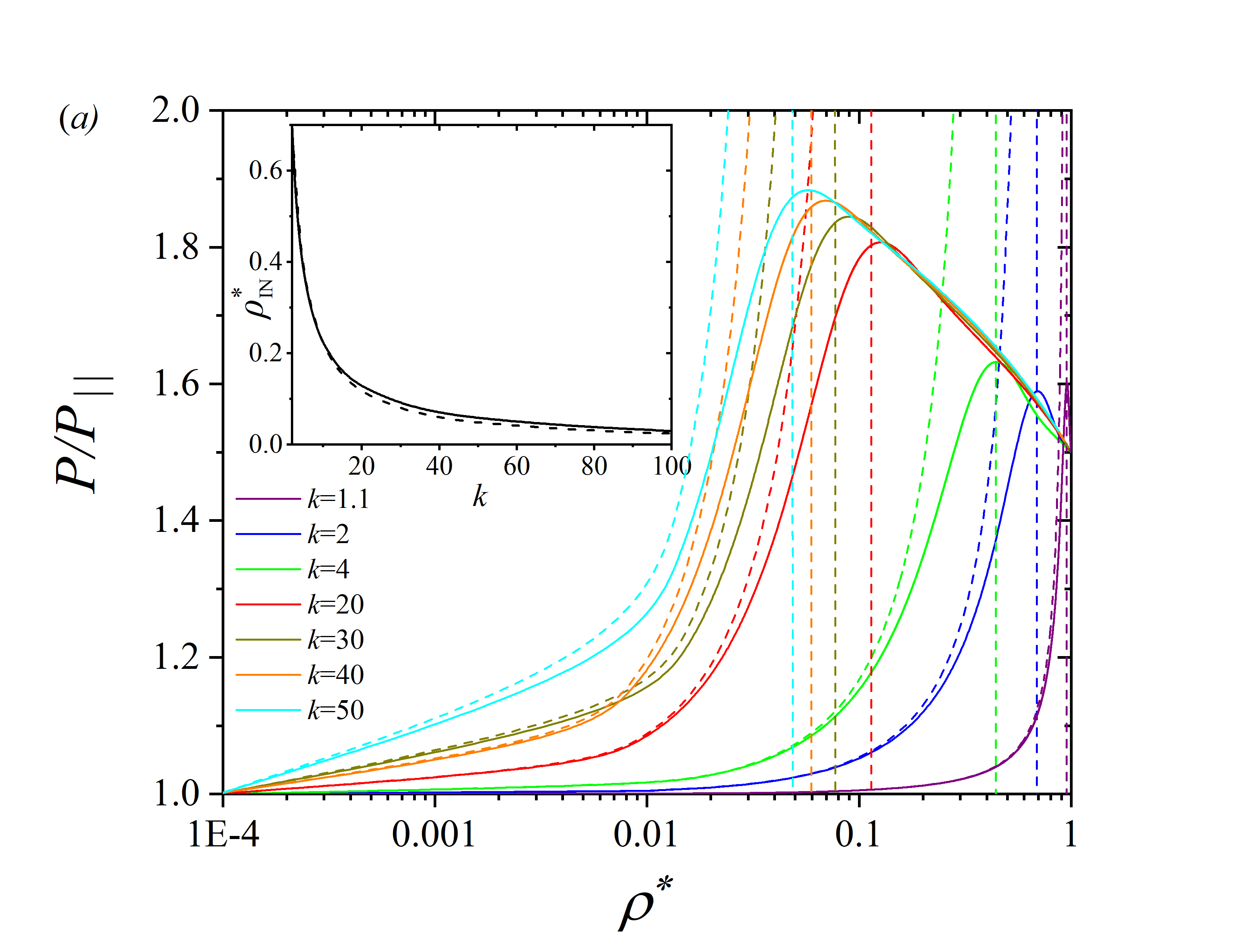}
        \includegraphics[width = 0.94\columnwidth,trim=1cm 0.5cm 3cm 2cm,clip ]{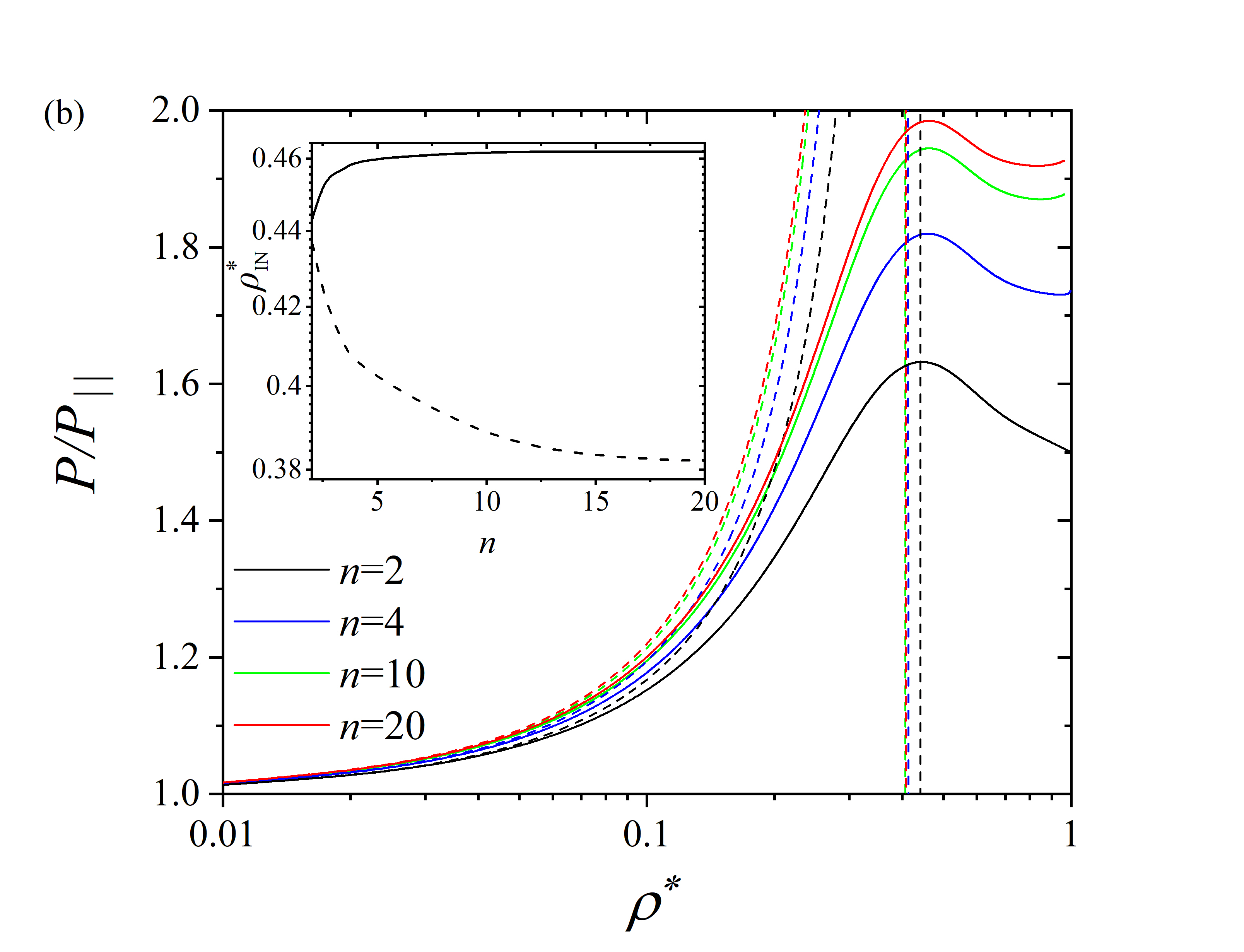}
        \includegraphics[width = 0.94\columnwidth,trim=1cm 0.5cm 3cm 2cm,clip ]{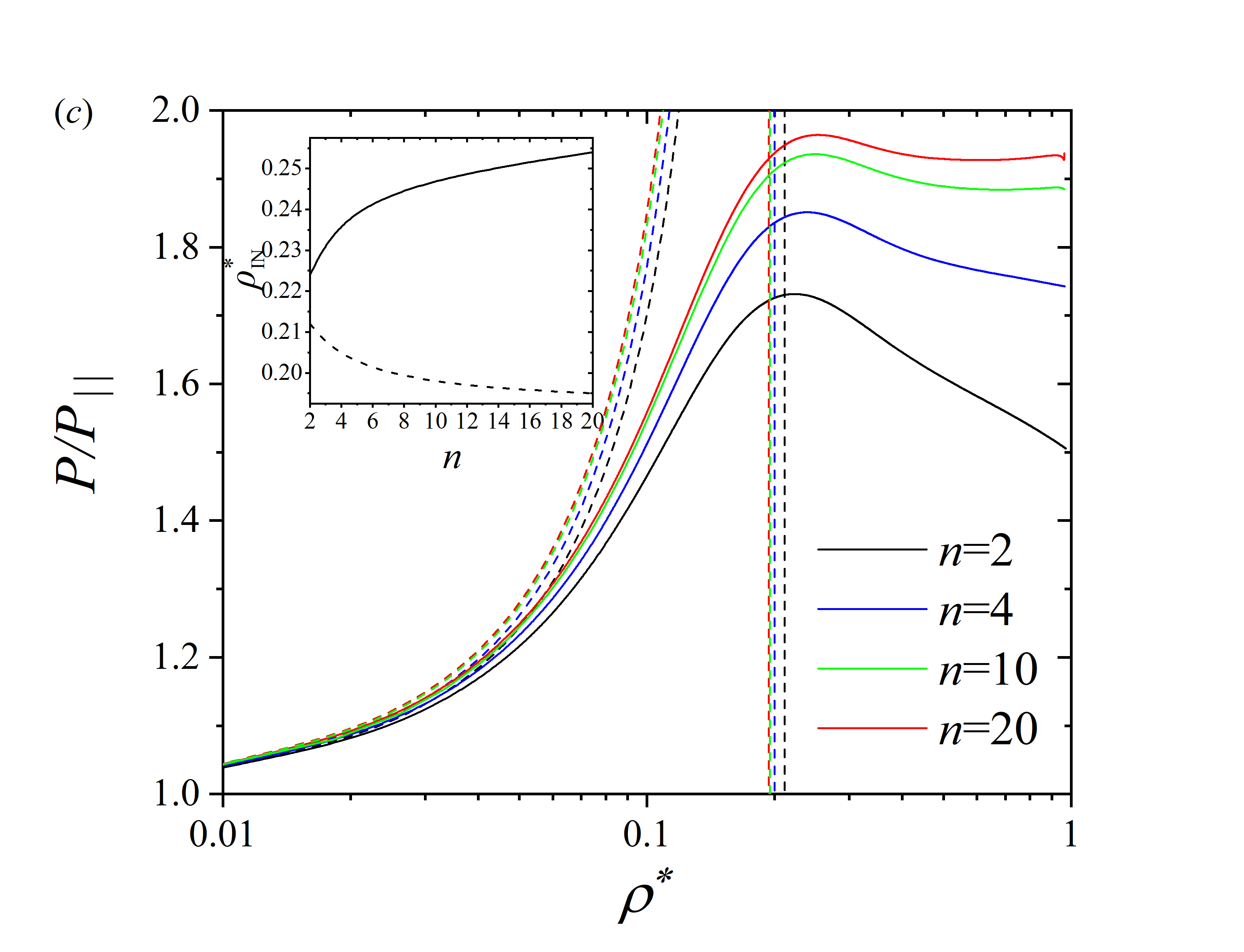}
     \caption{The pressure ratio $P/\phd$ as function of
    $\rho^*=\rho d$ for (a) hard ellipses ($n = 2$) and varying aspect
    ratio $k$, and for varying deformation parameter $n$ for (b) $k=4$
    and (c) $k=10$. Solid lines are exact results from TOM, dashed
    lines show $\piso/\phd$\ . Insets show the two approximations for
    the density of quasi-isotropic--nematic structural change (full lines:
    $\rmax$, dashed lines $\riso$, the latter is also marked by the
    dashed vertical lines in the main graphs). }
    \label{fig:pressure}
\end{figure}

In Fig.~\ref{fig:pressure} the (numerically) exact pressure ratio
$P/\phd$ is shown in three cases: the deformation parameter $n=2$ is
fixed (ellipses) and the aspect ratio $k$ is varying in
Fig.~\ref{fig:pressure}(a), while $n$ is varying and $k=4$ and $k=10$
is fixed in Fig.~\ref{fig:pressure}(b) and Fig.~\ref{fig:pressure}(c),
respectively.  We can see that the pressure ratio is always greater
than 1, and a general feature for all cases is the appearance of a
maximum in $P/\phd$ at some intermediate density $\rmax$. The pressure
ratio at the close-packing limit ($\rho d\rightarrow 1$) is
$\alpha=2-1/n$ (independent of $k$), which determines the first
close-packing exponent, see Eq.~(\ref{eq:alpha}). This can be shown exactly
using an appropriate expansion of the contact distance in
Eq.~\ref{eq:eigenvalue}, see also the discussion below in
Sec.~\ref{sec:discussion}.  An \textit{ansatz} for the pressure,
designed for quasi-isotropic states ($f(\varphi)\approx 1/\pi$), is
given by a modified Tonks equation of state
\begin{equation}
\piso=\frac{\kt \rho}{1-\rho\langle\sigma\rangle}\;, 
 \label{eq:piso}
\end{equation}
where the average contact distance
$\langle\sigma\rangle=\frac{1}{\pi^{2}} \int_{-\pi / 2}^{\pi/2} \int_{-\pi/2}^{\pi /2} \sigma\left(\varphi_{1}, \varphi_{2}\right) d \varphi_{1}d \varphi_{2}$ can be interpreted as an effective diameter for the
superellipses.  $\piso$ diverges at $\riso=1/\langle\sigma\rangle$,
which indicates a stability limit of the isotropic state.
In Fig.~\ref{fig:pressure}, $\piso/\phd$ is shown with dashed lines.
For $n=2$ and varying $k$, $\rmax$ nearly coincides with this
stability limit $\riso$ (see inset in Fig.~\ref{fig:pressure}(a))
which suggests the identification of $\rmax$ with a marker for the
structural change from a quasi-isotropic to the nematic state.  For fixed $k$,
$\rmax(n)$ and $\riso(n)$ show opposite behavior with increasing $n$,
although the absolute values are still similar (see insets in
Fig.~\ref{fig:pressure}(b) and (c)).  The decrease of $\rmax$ with
increasing $k$ and decreasing $n$ is in line with the observation in
2D and 3D that more anisotropic and less edgy rods are more prone to
form a nematic
phase~\cite{CampMasonAllenKhareKofke1996,BolhuisFrenkel1997,BatesFrenkel2000,LopesRomanoGreletFrancoGiacometti2021,BautistaCarbajalOdriozola2014,Dertli-Speck_PhysRevRes_2025}. {
These results agree with our finding for the orientational
distribution function 
(i.e. increasing $k$ supports nematic ordering while increasing $n$ 
supresses it). 
Consequently, the quasi-isotropic to nematic structural change is
shifted to 
higher densities with increasing $n$.}

\begin{figure}[ht]
   \centering
   \includegraphics[width=0.94\columnwidth,trim=1.5cm 0.1cm 3.2cm 2cm,clip]{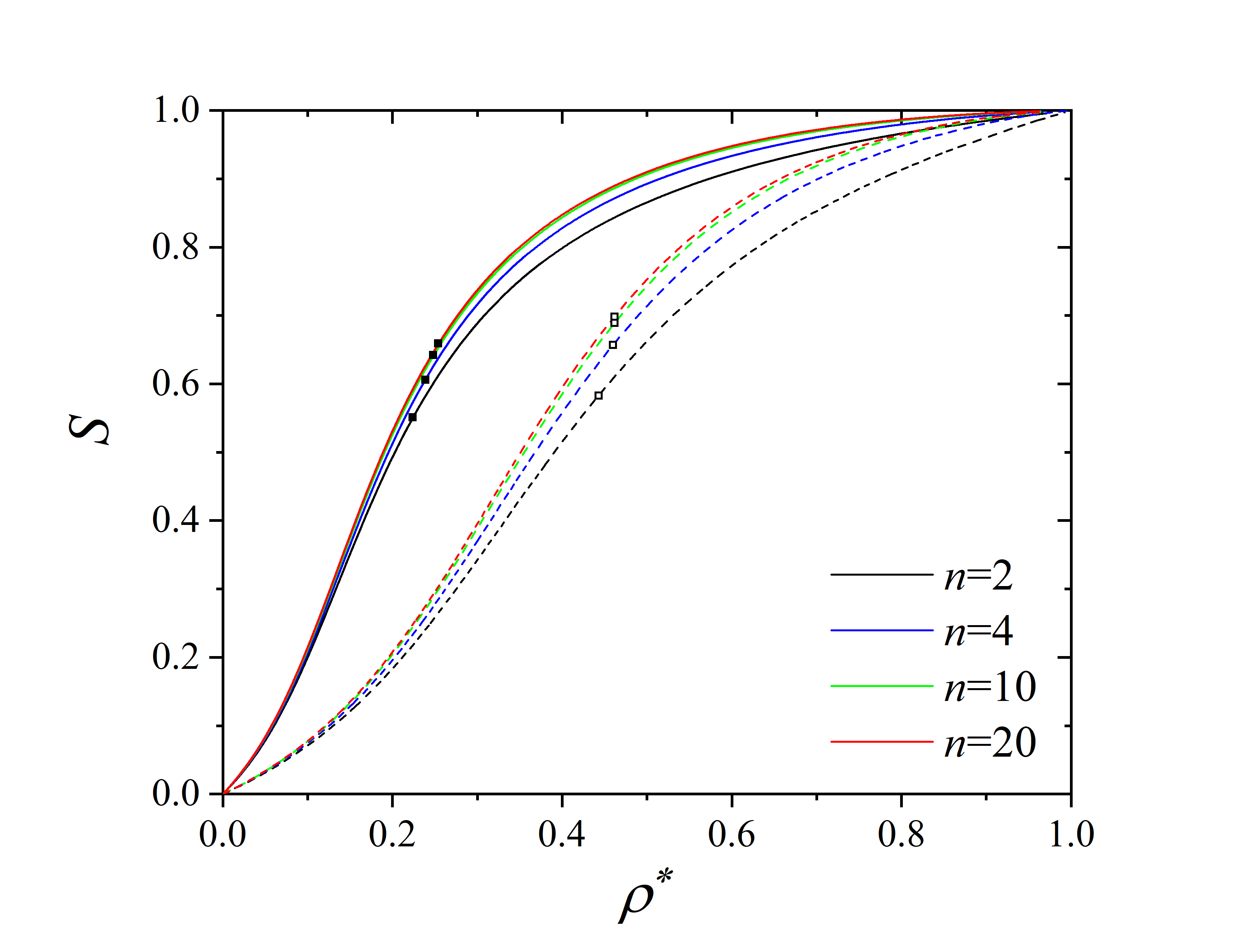}
   \caption{
    The nematic order parameter $S$ as a function of $\rho^*=\rho d$ for
    various values of \textit{n} (solid lines: $k=10$, dashed lines:
    $k=4$). Symbols indicate $S(\rmax)$.
    }
   \label{fig:order}
\end{figure}
Fig.~\ref{fig:order} shows the nematic order parameter $S$ as a
function of density $\rho$ for $k=4,\;10$ and various $n$. As expected
the change from a quasi-isotropic state $S\approx 0$ at low densities
to a nematic one ($S \lesssim 1$) near $\rho\sigma=1$ is
continuous. The order parameter at $\rmax$ (symbols in
Fig.~\ref{fig:order}) is at some intermediate value between 0.55 and
0.7, which further supports the identification of the pressure
ratio maximum as the indicator for structural change.  We refer also
to the discussion in Ref.~\cite{MizaniOettelGurinaVarga2024} on the
characteristics of the structural change for the case of superdisks
($k=1$).

\begin{figure}[ht]
   \centering
   \includegraphics[width=0.94\columnwidth,trim=1.cm 0.1cm 3.2cm 2cm,clip]{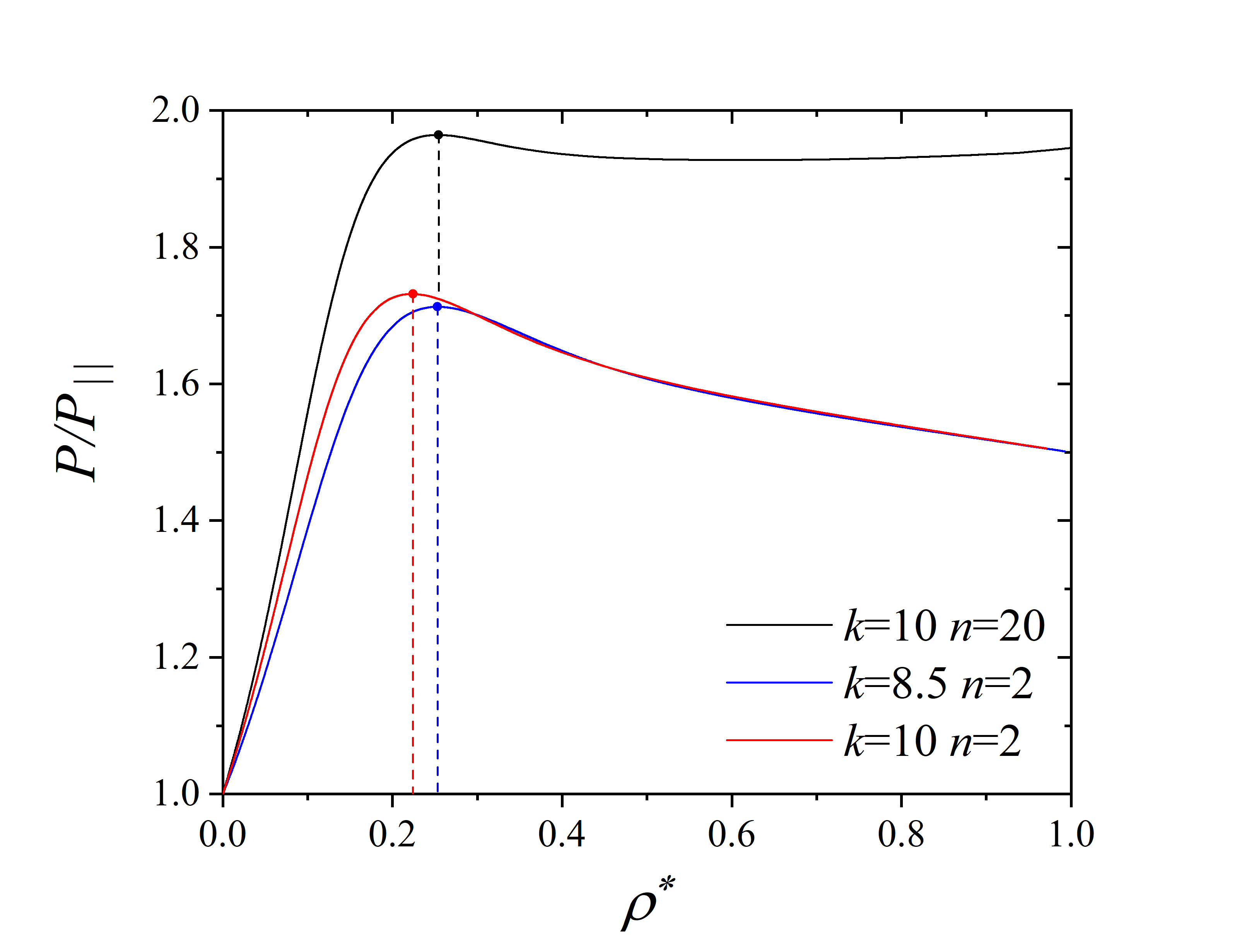}
   \caption{$P/\phd$ as a function of density for two values of  $k$ ($k=8.5$ and 10) and $n$ ($n=2$ and $n=20$). The vertical dashed lines mark the maximum of $P/\phd$ which approximately separates the quasi-isotropic and nematic structures.
   }
   \label{fig:p--rho_new} 
\end{figure}
{
An illustrative example for the effect of varying $k$ and $n$ on the peak in  $P/\phd$ is shown in
Fig.~\ref{fig:p--rho_new}. Taking ellipses ($n=2$) and increasing $k$ from 8.5 to 10, the peak is shifted to lower densities, i.e. nematic order is
stabilized with increasing $k$, which is a common feature of IN phase
transition observed in two- and three
dimensions~\cite{CuestaFrenkel1990,BatesFrenkel2000,Dertli-Speck_PhysRevRes_2025,Kike-Yuri-Luis_JPhysCondMat_2014}.  
On the other hand, for $k=8.5$, increasing now $n$ to 20 (near-rectangles) shifts the peak back to higher densities such that $\rmax$ is the same as in the ellipse, $k=8.5$ system. 
Thus, ellipses favor nematic ordering over rectangles. This feature is also observed in
two dimensions, where an elliptical shape is the most effective for 
nematic ordering~\cite{BautistaCarbajalOdriozola2014}. In the case of
hard rectangles ($k>1$ and $n\rightarrow\infty$), tetratic
ordering competes with nematic ordering, which results in a stable
tetratic phase for $k<8$~\cite{Dertli-Speck_PhysRevRes_2025}.  }

\begin{figure}[ht]
   \centering
   \includegraphics[width = 0.95\columnwidth,trim=0.05cm 0.4cm 5cm 2.5cm,clip]{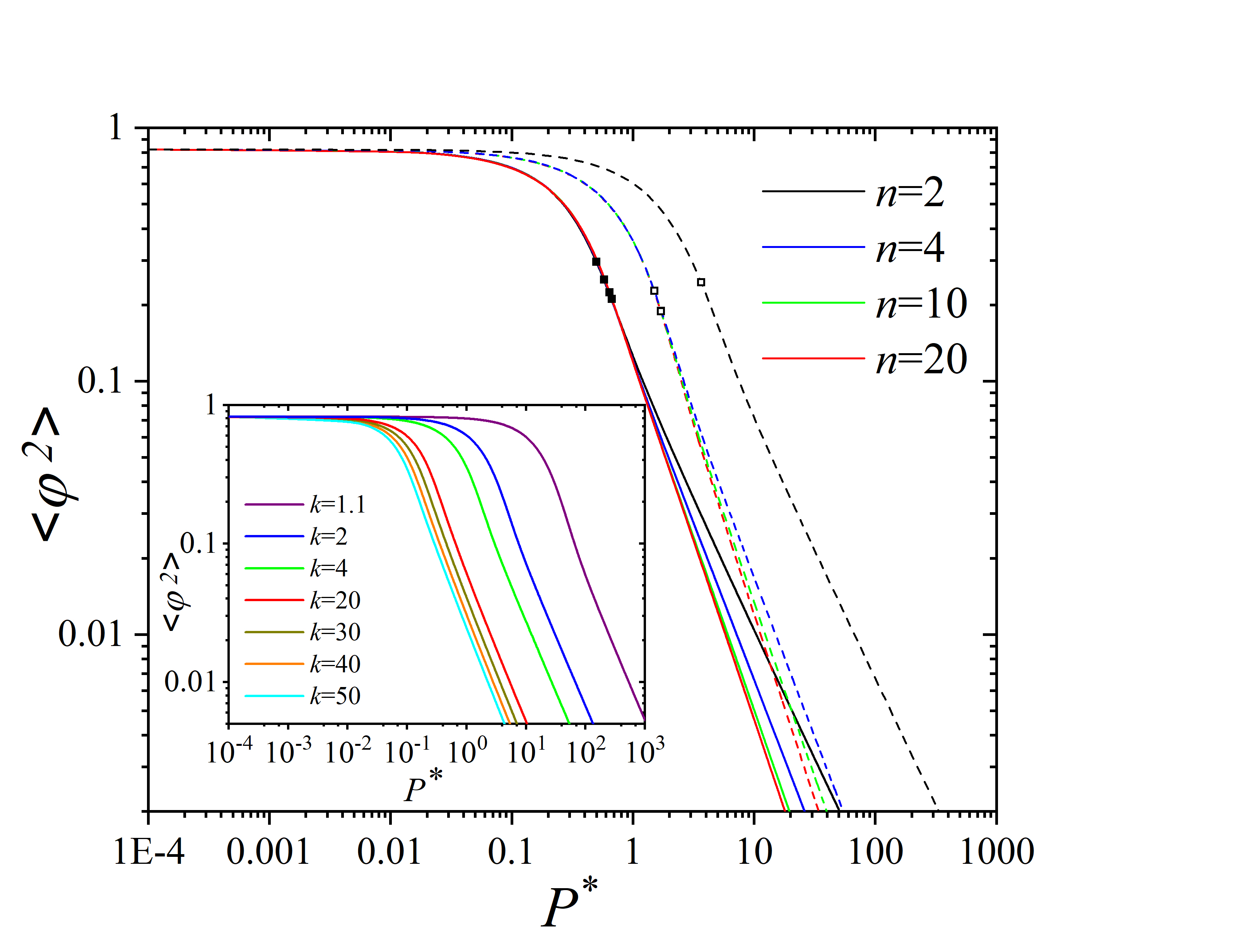}
   \caption{
    Average angular fluctuation as a function of pressure $P^*=\beta P d$
    for various values of the deformation parameter $n$ (symbols
    indicate the corresponding value at $P(\rmax)$).  Dashed lines are
    for $k=4$ and solid lines for $k=10$. The inset belongs to hard
    ellipses ($n=2$) with various aspect ratios $k$.
    }
   \label{fig:angfluc}
\end{figure}

The average angular fluctuations $\langle \varphi^2\rangle$ as
function of pressure are shown in Fig.~\ref{fig:angfluc} for two
aspect ratios $k=4$, 10 and various $n$ (main plot) and $n=2$
(ellipses) and various $k$ (inset). $\langle \varphi^2\rangle$ changes
from a constant at low density and pressure through a transition
region (location roughly defined by $P(\rmax)$) to the power-law
behavior $\propto P^\beta$ at high pressure. Numerically we find that
the close-packing exponent $\beta$ is independent of $k$ (see in
particular the inset of Fig.~\ref{fig:angfluc}) and consistent with
the simple expression $\beta=1 / n-3 / 2$.

\begin{figure}[ht]
   \centering
   \includegraphics[width = 0.95\columnwidth,trim=2.5cm 0.3cm 3cm 2.1cm,clip]{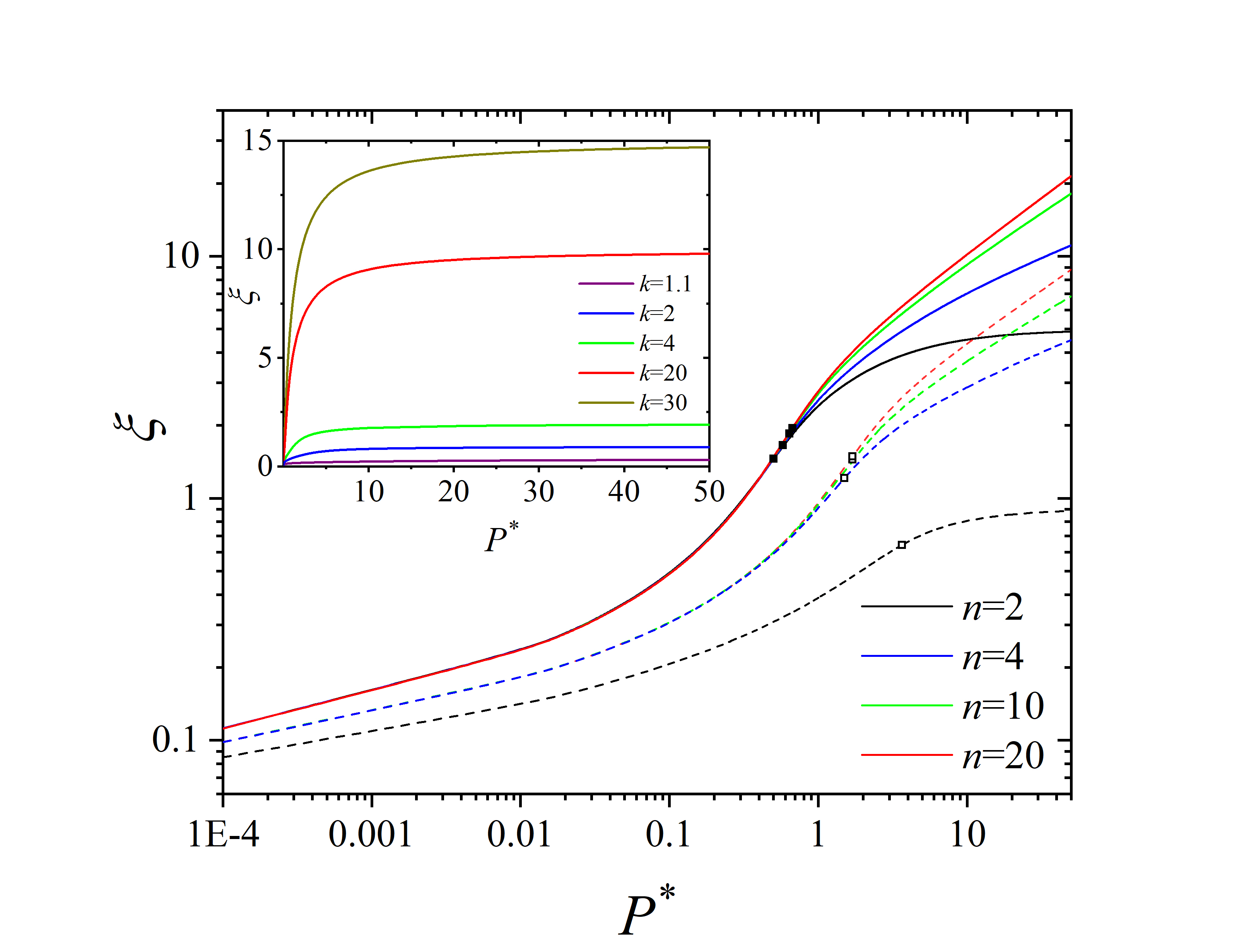}
   \caption{
    Orientational correlation length $\xi$ as function of $P^*=\beta P d$
    for various $n$ (symbols indicate $\xi$ at $P(\rmax)$).  Dashed
    lines are for $k=4$ and solid lines for $k=10$. The inset belongs
    to hard ellipses ($n=2$) with various aspect ratios $k$.
    }
   \label{fig:xi}
\end{figure}

The pressure dependence of the orientational correlation length
($\xi$) is shown in Fig.~\ref{fig:xi}.  As before, the main panel is
for the two aspect ratios $k=4$, 10 and various $n$ while the inset is
for $n=2$ (ellipses) and various $k$. The large pressure regime is
consistent with $\xi \propto P^\gamma$ and a value $\gamma=1 / 2-1 /
n$ for the close-packing exponent.  This implies a constant,
asymptotic correlation length for ellipses ($n=2$).

\begin{figure}[t]
   \centering
   \includegraphics[width = 0.95\columnwidth,trim=2cm 1cm 3cm 1cm,clip]{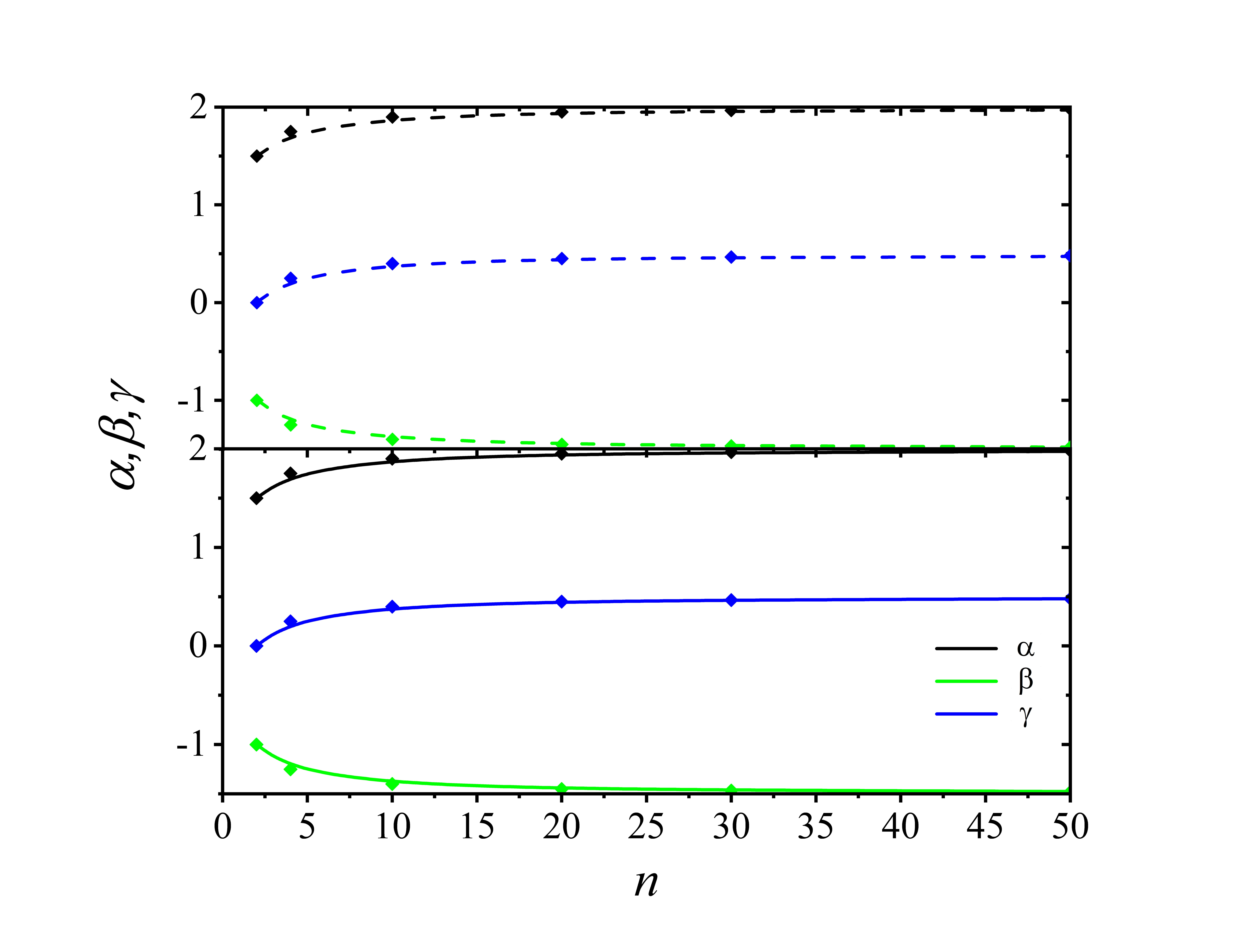}
	\caption{
        $\alpha$, $\beta$, and $\gamma$ exponents as functions of the
	deformation parameter (\textit{n}). The symbols represent
	numerical values obtained from TOM, while the dashed and
	continuous lines are the curves $\alpha=2-1 / n, \beta=1 / n-3
	/ 2$ and $\gamma=1 / 2-1 / n$. The upper panel belongs to $k=
	4$, while the lower one to $k= 10$.  } \label{fig:crit}
\end{figure}

Finally, Fig.~\ref{fig:crit} examines the validity of the proposed
formulas for the close-packing exponents: $\alpha=2-1 / n, \beta=1 /
n-3 / 2$ and $\gamma=1 / 2-1 / n$.  The numerical data nearly
perfectly match those.  In our previous
study \cite{MizaniOettelGurinaVarga2024}, the above formulas were
found to be valid for hard superdisk $(k=1)$. The present results show
that the close-packing exponents do not depend on the aspect ratio
$k$.  {This implies that the close
packing behaviour of superellipses and superdisks becomes identical
even if the structures are not the same (nematic for $k>1$ and tetratic
for $k=1$).} Notably, the analytical expressions for these parameters reveal
a remarkable shape independent universal rule for the combined
quantities $\alpha+\beta=1/2$, $\beta+\gamma=-1$ and
$\alpha+2 \beta+\gamma=-1/2$ (independent of $n$ and
$k$).  The results strongly suggest that the universal rules also hold
for q1D fluids with rod-like shapes other than superelliptical ones.

\section{Discussion and Conclusion}
\label{sec:discussion}

In this study, we have examined the orientational ordering behavior
{(induced by the deformation parameter $n$ and the aspect ratio $k$)}
and close packing properties of q1D fluid of hard superellipses.
{
While $n>2$ alone gives rise to orientational ordering with fourfold symmetry (tetratic ordering), $k>1$ produces ordering with two-fold symmetry (nematic ordering). In the case of superellipses ($k>1$ and $n\geq2$) the ordering is always nematic, but some similarities with tetratic ordering are present for $1<k<2$ and $n>2$, as seen by the double-peaked orientational distribution function favoring both parallel and perpendicular ordering. However, this intermediate orientational ordering is present only at low densities, while the distribution function is always single peaked at high densities. 

At low densities, superellipses form a quasi-isotropic structure (only weakly ordered in the orientation), while the structure becomes perfectly ordered nematic at the close-packing density. The structural change from quasi-isotropic to perfect nematic is not simple with increasing pressure (density), because orientational entropy is maximal in the isotropic structure, while packing entropy is maximal for perfect nematic ordering. The competition of these two entropies produces a non-monotonic change in the order parameter, which has a convex shape at low densities and becomes concave at high densities (see Fig.~\ref{fig:order}). This shows that the orientational entropy is dominant at low densities, while the packing one at high densities. Note that the competition between orientational and packing entropies is also present in higher dimensions, where they produce isotropic--nematic phase transition~\cite{Kike-Yuri-Luis_JPhysCondMat_2014}. However, this competition produces only a structural change in quasi-one dimension.

In the absence of a phase transition, we have found that the ratio $P/\phd$ with its emerging maximum is a very efficient marker to distinguish the low density quasi-isotropic structure from the high density nematic one. The effects of varying $k$ and $n$ are opposite for locating the change from quasi-isotropic to nematic structures, because $k$ supports the nematic ordering, while $n$ weakens it. This result is due to the fact that that increasing $n$ produces sharper edges, which is crucial for tetratic ordering, while increasing $k$ makes the particle more elongated and stabilizes the nematic ordering. Note that similar trends was observed in higher dimensions, where the sharp edges (high deformation parameter) destabilizes the nematic phase. For example, an elliptical shape is more suitable for nematic ordering than a rectangular one.
}

{The competition between $n$ and $k$ becomes weaker in the strongly ordered nematic region, which manifests very clearly in the vicinity of close-packing density. We have observed that the close packing exponents ($\alpha$, $\beta$ and $\gamma$) are independent of $k$, i.e. the close packing behaviour of hard superdisks and that of hard superellipses are the same, although superdisks form perfect tetratic and superellipses nematic structures.}
To better understand this unusual behaviour, we examine the contact
distance between two superellipses in the special case when the
particles are almost parallel and align along the $y$-axis (as this is
the relevant configuration near close packing, see
Fig.~\ref{fig:geometry}).  Taylor-expanding around
$\varphi_{1}=\varphi_{2}=0$, the leading terms are shown to be:
\begin{eqnarray}
    \label{eq:sigana}
    	\sigma\left(\varphi_{1}, \varphi_{2}\right)&=&a\left(1+\frac{\varphi_{p}^{2}-\varphi_{m}^{2}}{2}-\right. \\
        && \left. \frac{1}{n k^{n}}\left|\varphi_{p}\right|^{n}+\frac{n-1}{n} k^{\frac{n}{n-1}}\left|\varphi_{m}\right|^{\frac{n}{n-1}}\right) \;, \nonumber
\end{eqnarray}
where $\varphi_{p}=\frac{\varphi_{1}+\varphi_{2}}{2}$ and
$\varphi_{m}=\frac{\varphi_{1}-\varphi_{2}}{2}$.  In this equation,
$n$ appears both in the prefactor and exponent of the $\varphi_{p}$
and $\varphi_{m}$ terms, while $k$ appears in the prefactors only.
Using Eq.~(\ref{eq:sigana}) in the eigenvalue problem
(\ref{eq:eigenvalue}) gives eigenvalues $\lambda_0$, $\lambda_1$ as
well as eigenfunctions $\psi_0$, $\psi_1$ which are independent of $k$
for $p\rightarrow\infty$.  The key term of Eq.~(\ref{eq:sigana}) is
$\left|\varphi_{m}\right|^{\frac{n}{n-1}}$, which determines the
``cost'' of angular fluctuations in the contact distance. As a result,
$\alpha, \beta$ and $\gamma$ depend only on $n$.  Note here that if
the close packing configuration is degenerate in the orientation, the
above rules for close-packing exponents are not
valid \cite{GurinaMizaniVarga2024}.

A remarkable consequence of the aspect ratio independence of the close-packing exponents
{is that $\alpha+\beta=1/2$, $\alpha+2\beta+\gamma=-1/2$ and $\beta+\gamma=-1$ are general rules for hard superellipses, i.e. the close-packing exponents of many shapes including the limiting square, ellipse and rectangle shapes obey the same rules. However, the validity of these rules must be examined for other classes of shapes in order to draw a definitive conclusion on their overall validity. It would be also interesting to study 3D hard bodies with full 3D rotational freedom to find similar behaviour in the close-packing exponent.
}

\section*{Acknowledgements}
S. M. acknowledges the financial support of Humboldt
Foundation. P. G. and S. V. gratefully acknowledge the financial
support of the National Research, Development, and Innovation Office -
NKFIH K137720 and the TKP2021-NKTA-21.

\bibliography{paper-revised}

\end{document}